\documentstyle[preprint,amsfonts,amsmath,prd,aps,
12pt,epsf,epsfig]{revtex}
\begin{document}
\preprint{
\vbox{
\halign{&##\hfil\cr
         & HUPDQ-0202\cr
         & hep-ph/0206093\cr
         & June 2002 \cr
         & \cr
         & \cr
}}}
\vskip 4mm
\title{$J/\psi$ Pair Production at the Tevatron}
\author{Cong-Feng Qiao\footnote{JSPS Research Fellow. 
E-mail: qiao@theo.phys.sci.hiroshima-u.ac.jp}}
\vskip 1pt
\address{Department of Physics, Faculty of Science,\\
Hiroshima University, Higashi-Hiroshima 739-8526, Japan}
\maketitle
\vskip 2.5cm
\centerline{\bf Abstract}
\vskip 7mm
\begin{minipage}{145mm}
The double $J/\psi$ production in $p\,\bar{p}$ collision 
is revisited. It is found that the $J/\psi$ pair production rate 
at leading order with $p_T > 4$ GeV in conventional scheme is larger
than the rate derived from color-octet mechanism by a factor of five. 
As the double $J/\psi$ production is shown to be attainable 
with data yet collected at the Tevatron detectors, experimental study 
on it would be helpful to clarify the validity of the conventional 
color-singlet description for heavy quarkonium production.
\vskip 9pt

\noindent
PACS Number(s){12.38.Bx, 13.85.Ni, 14.40.Lb}

\end{minipage}
\vfill \eject
Quarkonium production and decays have long been taken as an ideal 
means to investigate the nature of QCD and other new 
phenomena. Hence, to establish a proper theory which can 
precisely describe heavy quarkonium production and decays 
is very necessary. A novel effective theory, 
non-relativistic QCD(NRQCD) \cite{nrqcd}, is possibly one 
theory to this aim which is formulated from the first 
principles. However, to make a precise prediction for 
quarkonium production, with only the NRQCD is not enough 
at least for now, since the magnitude of the 
non-perturbative parameters in the theory are still 
unknown. Therefore, generally speaking theoretical 
calculation results may depend on the values of the input 
non-perturbative parameters, although NRQCD can give 
out the relative weights of these parameters in order of 
$v^2$ from its "velocity scaling rules". Up to now, on one 
hand the color-octet \cite{com} mechanism still stands as 
the most plausible proposal in explaining the large transverse 
momentum $\psi(\psi')$ production "anomaly" \cite{cdf}; on the 
other hand, it encounters some difficulties in 
confronting other phenomena \cite{rothstein}. Especially,
a recent research \cite{qcf1} shows that previous studies 
of the color-octet contributions to high-$p_T$ 
$J/\psi(\psi')$ production were overestimated, 
although not much, which may somehow 
bring new light to the understanding of heavy quarkonium, 
especially for charmonium, production mechanism. All in all, 
to what degree the color-octet mechanism plays the role 
in quarkonium production is still not clear and an 
interesting question. 

During past decade, lots of efforts 
have been made in proposing approaches to find definite 
color-octet signals. Unfortunately, the nature appears to 
be more elusive than expected and experiments still give 
no conclusive answer to this demand yet. One interesting 
point is that once there is a proposal highlighting a octet 
process, not very long people find another competing 
process in singlet, which brings difficulties to 
experimenters in distinguishing between color-octet and singlet. 
For instance, in electron-position scattering Ref. \cite{braatenc} 
to Refs. \cite{changqw}, and Ref. \cite{qcf2} to Ref. 
\cite{mamp} in direct photon-photon collision.

In explaining the high-$p_T$ $\psi$ surplus production 
discovered by CDF group \cite{cdf} at the Fermilab 
Tevatron, the color-octet scheme tells us that the dominant source of 
charmonium production at large transverse momentum comes from 
the produced hard gluon followed by its fragmentation into an 
intermediate color-octet $c\bar{c}$ state, which eventually evolves 
into quarkonium non-perturbatively. Applying this idea to the 
double-quarkonium production, Barger {\it et al}. \cite{barger} 
studied the quarkonium pair production via the double-gluon 
fragmentation and found that this scheme for double $J/\psi$ production
gives a result which is measurable with the accumulated data at
the Fermilab Tevatron detectors. And hence, claimed the measurement
of this process in experiment would give a test on the color-octet
mechanism. At sufficient high transverse momentum this treatment 
might be true, however, at moderate and relatively low $p_T$ one 
should take great care. In the present work we show that it is 
really the case for double-$J/\psi$ production at even not very 
low transverse momentum.

In color-singlet model, the partonic subprocess starts at 
order $\alpha_s^4$. The lowest order processes include 
$g + g \rightarrow J/\psi + J/\psi$ 
and $q + \bar{q} \rightarrow J/\psi + J/\psi$. 
Since the latter, the quark annihilation process, 
reasonably gives less contribution at Tevatron energy,
in this brief report, we restrict our calculation only to 
the gluon-gluon fusion one, as shown in figure 1. 
This parton process is similar to the case in 
photon-photon scattering discussed in Ref.\cite{qcf2}, however, 
the extension to the present discussion is not trivial. The QCD 
non-Abelian nature makes more possible channels involve in, i.e., 
the third topological group shown in the Figure.

The differential cross section for quarkonium pair hadroproducion is 
given by
\begin{eqnarray}
\label{eq:1}
\frac{d\sigma}{d p_T} (p \bar{p} \rightarrow 2J/\psi + X)  = 
\sum_{a,b}\int dx_a dy_1 f_{a/p}(x_a) f_{b/\bar{p}}(x_b) 
\frac{4 p_T x_a x_b}{2 x_a - \bar{x}_T e^{y_1}} \frac{d\hat{\sigma}}
{d{t}} (a + b \rightarrow 2J/\psi + X)\;,
\end{eqnarray}
where $f_{a/p}$ and $f_{b/\bar{p}}$ denote the parton densities;
$t$, as well as $s$ and $u$ appearing later, is Mandelstam variable
of the parton level; $y_1(y_2)$ is of the rapidity of produced 
$J/\psi$; $\bar{x}_T \equiv \frac{2 m_T}{\sqrt{\bf S}}$ with $m_T = 
\sqrt{m^2 + p_T^2}$. Here, the capital $\sqrt{\bf S}$ denotes 
the incident beams total energy, and taken to be 1.8 TeV of the 
Tevatron in our subsequent numerical calculations. 

The parton scattering differential cross section of gluon-gluon 
to $J/\psi$ pair can be calculated by the standard way 
straightforwardly. It is
\begin{eqnarray}
\label{diff}
\frac{d \hat{\sigma}}{d {t}} &=& \frac{16\ \alpha_s^4 \ \pi \ |R(0)|^4}
{81 m^2 s^8 (m^2 - t)^4  (m^2 - u)^4}
\left[2680 m^{24} - 14984 m^{22} t + 31406 m^{20} t^2 - 31824 m^{18} 
t^3
\right. 
\nonumber\\
&+& 
   17668 m^{16} t^4 - 7172 m^{14} t^5 + 2956 m^{12} t^6 - 794 m^{10} t^7 + 
   47 m^8 t^8 + 20 m^6 t^9 + m^4 t^{10} \nonumber\\
&-& 
   14984 m^{22} u + 89948 m^{20} t u - 202576 m^{18} t^2 u + 
   228560 m^{16} t^3 u - 153360 m^{14} t^4 u \nonumber\\
&+& 
   76406 m^{12} t^5 u - 30782 m^{10} t^6 u + 7642 m^8 t^7 u - 822 m^6 t^8 u - 
   66 m^4 t^9 u + 31406 m^{20} u^2 \nonumber\\
&-&
   202576 m^{18} t u^2 + 470856 m^{16} t^2 u^2 - 536476 m^{14} t^3 u^2 + 
   361624 m^{12} t^4 u^2 - 182454 m^{10} t^5 u^2 \nonumber\\
&+& 
   73146 m^8 t^6 u^2 - 17902 m^6 t^7 u^2 + 2469 m^4 t^8 u^2 + 
   36 m^2 t^9 u^2 - 31824 m^{18} u^3 + 228560 m^{16} t u^3 \nonumber\\
&-& 
   536476 m^{14} t^2 u^3 + 571900 m^{12} t^3 u^3 - 335186 m^{10} t^4 u^3 + 
   150334 m^8 t^5 u^3 - 58126 m^6 t^6 u^3 \nonumber\\
&+& 12874 m^4 t^7 u^3 - 2344 m^2 t^8 u^3 + 17668 m^{16} u^4 - 
   153360 m^{14} t u^4 + 361624 m^{12} t^2 u^4 \nonumber\\
&-& 335186 m^{10} t^3 u^4 + 132502 m^8 t^4 u^4 - 35306 m^6 t^5 u^4 + 
   11928 m^4 t^6 u^4 - 148 m^2 t^7 u^4 + 698 t^8 u^4 \nonumber\\
&-& 
   7172 m^{14} u^5 + 76406 m^{12} t u^5 - 182454 m^{10} t^2 u^5 + 
   150334 m^8 t^3 u^5 - 35306 m^6 t^4 u^5 + 1164 m^4 t^5 u^5 \nonumber\\
&-& 
   1576 m^2 t^6 u^5 - 1816 t^7 u^5 + 2956 m^{12} u^6 - 30782 m^{10} t u^6 + 
   73146 m^8 t^2 u^6 - 58126 m^6 t^3 u^6 \nonumber\\
&+& 11928 m^4 t^4 u^6 - 1576 m^2 t^5 u^6 + 2748 t^6 u^6 - 794 m^{10} u^7 + 
  7642 m^8 t u^7 - 17902 m^6 t^2 u^7 \nonumber\\
&+& 12874 m^4 t^3 u^7 - 148 m^2 t^4 u^7 - 1816 t^5 u^7 + 47 m^8 u^8 - 
   822 m^6 t u^8 + 2469 m^4 t^2 u^8  \nonumber\\
&-& 2344 m^2 t^3 u^8
 + 698 t^4 u^8 + 20 m^6 u^9 - 66 m^4 t u^9 + 36 m^2 t^2 u^9 + m^4 u^{10}
\left.\right]\;,
\end{eqnarray}
where $m$ is the mass of charmonium, the $J/\psi$; $|R(0)|$ is the 
magnitude of its radial wavefunction at origin. In obtaining the above 
analytical expression, we start from general Feynman rules and project 
the Charm-anti-Charm pair into the S-wave vector charmonium state in 
color-singlet. To manipulate the trace and matrix-element-square of 
those tens of diagrams, the computer algebra system MATHEMATICA is 
employed with the help of the package FEYNCALC 
\cite{feyncalc}. 

The values of input parameters used in our numerical calculations are:
\begin{eqnarray}
m_c =  1.5\; \rm{GeV},\; |R(0)|^2 = 0.8\;\rm{GeV}^3\;.
\end{eqnarray}
Here, the non-relativistic relation $m = 2 m_c$ is adopted. 
The typical scale is set to be at $m_T$, and 
hence the strong coupling is running with transverse momentum.

With the above preparation, direct $J/\psi$ pair production cross 
section can be immediately obtained. Applying the pseudorapidity 
cuts on both produced charmonia, 
i.e. $|\eta(\psi_1)|, |\eta(\psi_2)| < 0.6$, the integrated
cross section 
$\sigma(p\bar{p} \rightarrow \psi_{\mu^+\mu^-}\psi_{\mu^+\mu^-})$
for $p_T(\psi) > 4$ GeV is 0.70 pb. Here, the notation
$\psi_{\mu^+\mu^-}$ means that the branching ratio of 
$B(\psi \rightarrow {\mu^+\mu^-}) =  0.0588$ of practical measuring 
mode to reconstruct the charmonium state is included.
For integrated luminosity of 100 $\rm{pb}^{-1}$ achieved in
the past run of the Tevatron, there should be about seventy 
$J/\psi$ pair events coming from the conventional scheme, 
which is about five times larger than what
obtained in Ref. \cite{barger}. In our numerical calculation, the
parton distribution of CTEQ5L \cite{cteq} is used, and both 
renormalization scale and factorization scale are evolved to the same 
point $m_T$. In the numerical calculation, the integration limits of 
$x_a$ in Eq. (\ref{eq:1}) are truncated in accordance with the 
pseudorapidity cuts in both produced charmonia.

The double-$J/\psi$ production cross section as function of transverse 
momentum $p_T$ is shown in Figure 2 in solid line. For comparison, the 
result from color-octet process discussed in Ref. \cite{barger} is 
presented in the same figure in dashed line. From the figure we see 
the conventional production scheme dominates over the octet one in 
relatively low-$p_T$ region, $p_T < 7$ GeV, where most of the 
charmonium pairs are produced. In principle, the color-octet 
mechanism, at least the importance of it, can be undisputedly 
checked in large transverse momentum area, however, unfortunately 
there is no enough available data for such purpose at the moment.

In above analysis, we take charmonium state, the $J/\psi$, as an 
object. Nevertheless, the results can be readily applied to some other 
quarkonium states, both charmonium and bottomonium. For example 
the $\psi'$ pair production rates in color-singlet model can be 
obtained by multiplying the constant
\begin{eqnarray}
\frac{|R'(0)|^4}{|R(0)|^4}\frac{B^2(\psi' \rightarrow {\mu^+\mu^-})}
{B^2(J/\psi \rightarrow {\mu^+\mu^-})} \approx
\frac{B^4(\psi' \rightarrow {\mu^+\mu^-})}
{B^4(J/\psi \rightarrow {\mu^+\mu^-})}  
\end{eqnarray}
and the $J/\psi$ results.

In conclusion, we have calculated the $J/\psi$ pair production rate 
at the Fermilab Tevatron with the conventional heavy quarkonium 
production treatment, the color-singlet model. The calculation is 
carried out at the leading order in strong coupling constant. Whereas, 
the higher order corrections can be taken to be small due to 
relatively high interacting scale we have considered; the relativistic 
correction, which is not always too small to be negligible in dealing 
with charmonium system, is also double suppressed since we are 
considering the pair production; and for the same reason, the higher 
excited states feeddown is well suppressed as well. From our analysis, 
the accumulated data at the Fermilab Tevatron detectors can give 
about one hundred of $J/\psi$ pair events with transverse momentum
$p_T > 4$ GeV, among them about eighty
percent is produced via the conventional quarkonium production
mechanism. This finding tells us that analyzing the data 
for this process is not a suitable game in chasing the color-octet 
signatures, but rather provides an ideal means to precisely check 
the conventional description for heavy quarkonium production.

\vskip 1.2cm
\centerline{\bf ACKNOWLEDGEMENTS}
\vskip 0.3cm
This work was supported by the Grant-in-Aid aid of JSPS committee. 
The author is grateful to the organizers of RIKEN school 2002, when 
this work was initiated.

\begin{figure}
\vskip -3cm
\epsfxsize=15 cm
\centerline{\epsffile{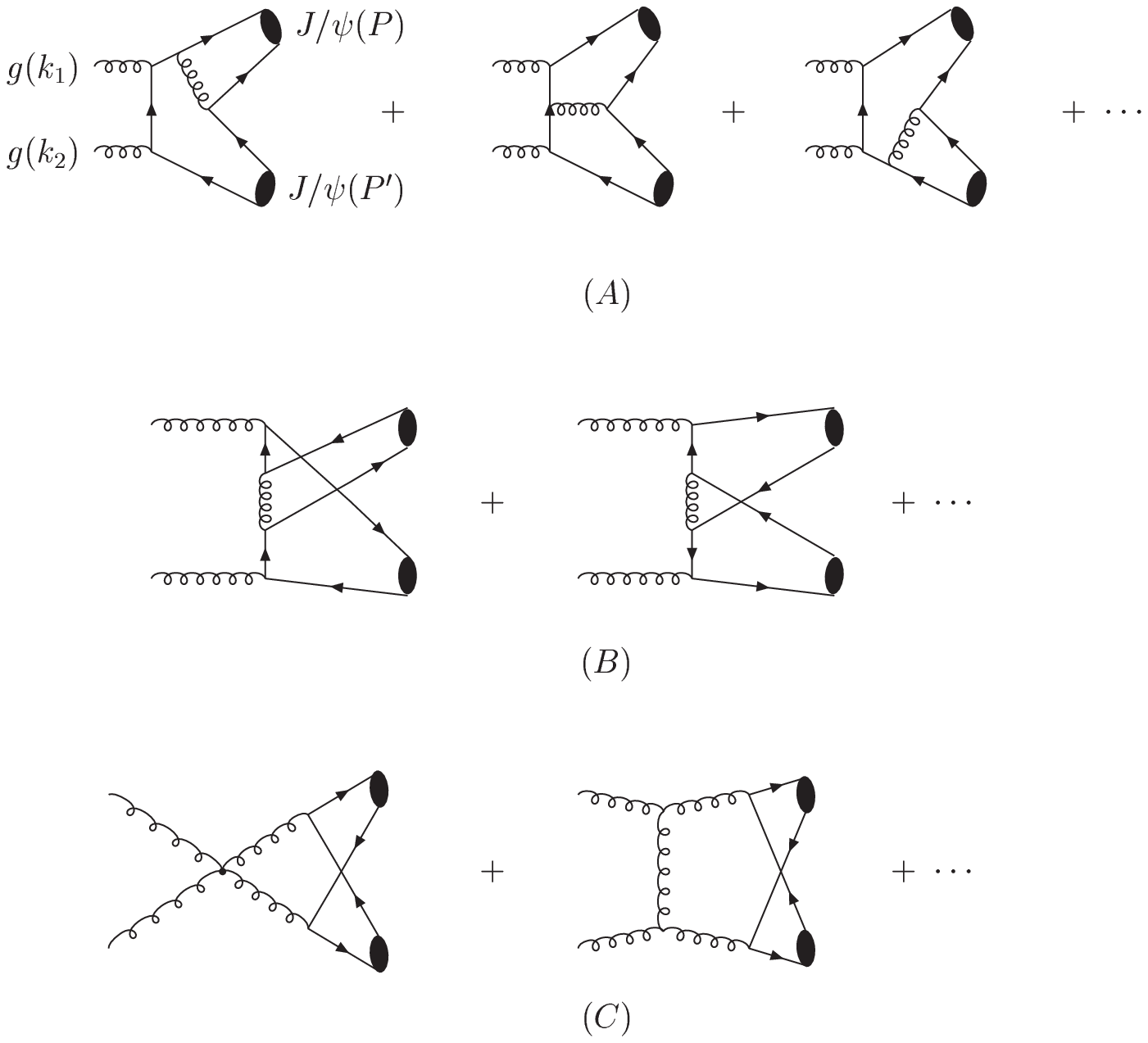}}
\vskip -7.2cm
 \caption[]{Typical Feynman diagrams of $J/\psi$ pair production in
$p\,\bar{p}$ collision at leading order.}
\label{graph1}
\end{figure}    

\begin{figure}[tbh]
\begin{center}
\vspace{-3cm}
\epsfig{file=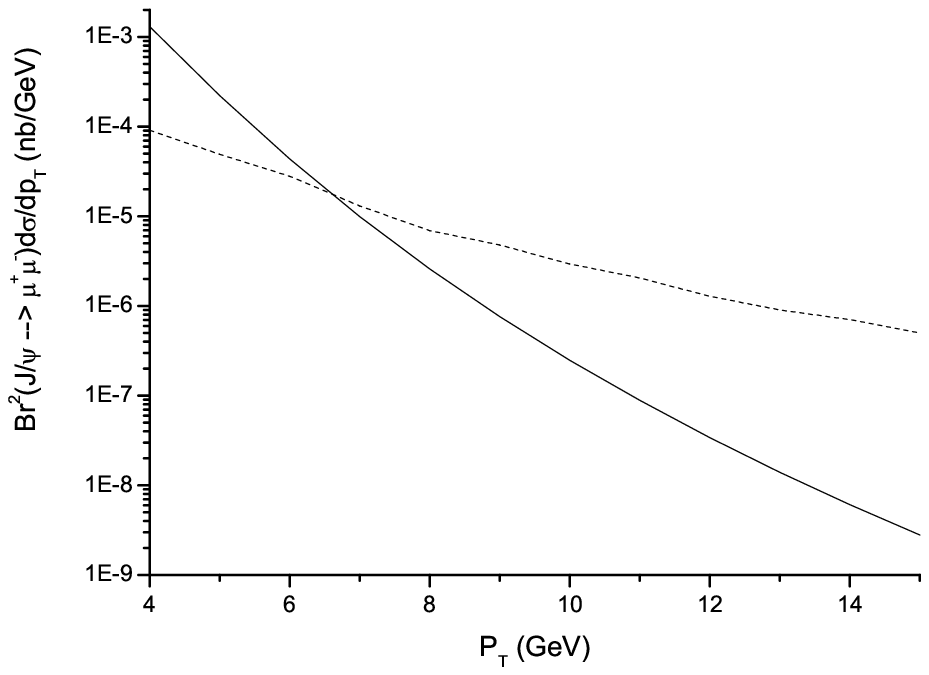,bbllx=90pt,bblly=300pt,bburx=230pt, 
bbury=420pt,width=4cm,height=2.5cm,clip=0}
  \end{center}
\vskip 6cm
  \caption[bt]{The differential cross-section of $J/\psi$ pair 
production versus $p_T$ at the Tevatron. Solid line comes from the 
Color-Singlet calculation of this paper; the dashed line from the 
Color-Octet calculation read from Ref.\cite{barger}.}
  \label{graph3}
\end{figure}    
\end{document}